\documentclass[aps,prl,onecolumn,superscriptaddress]{revtex4-1}
\usepackage{graphicx}
\usepackage{here}
\usepackage{xspace}
\usepackage{amsmath}
\usepackage{amssymb}
\usepackage{color}
\begin{document}
\title{Change of Fermi surface states related with two different $T_{\rm c}$-raising mechanisms in iron pnictide superconductor}

\author{A.~Takemori}
\affiliation{Department of Physics, Osaka University, Osaka 560-0043, Japan.}
\author{T.~Hajiri}
\affiliation{Department of Materials Physics, Graduate School of Engineering, Nagoya University, Nagoya 464-8603, Japan.}
\author{S.~Miyasaka}
\affiliation{Department of Physics, Osaka University, Osaka 560-0043, Japan.}
\author{Z.~H.~Tin}
\affiliation{Department of Physics, Osaka University, Osaka 560-0043, Japan.}
\author{T.~Adachi}
\affiliation{Department of Physics, Osaka University, Osaka 560-0043, Japan.}
\author{S.~Ideta}
\affiliation{UVSOR Facility, Institute for Molecular Science, Okazaki 444-8585, Japan.}
\affiliation{School of Physical Sciences, The Graduate University for Advanced Studies (SOKENDAI), Okazaki 444-8585, Japan.}
\author{K.~Tanaka}
\affiliation{UVSOR Facility, Institute for Molecular Science, Okazaki 444-8585, Japan.}
\affiliation{School of Physical Sciences, The Graduate University for Advanced Studies (SOKENDAI), Okazaki 444-8585, Japan.}
\author{M.~Matsunami}
\affiliation{UVSOR Facility, Institute for Molecular Science, Okazaki 444-8585, Japan.}
\affiliation{School of Physical Sciences, The Graduate University for Advanced Studies (SOKENDAI), Okazaki 444-8585, Japan.}
\author{S.~Tajima}
\affiliation{Department of Physics, Osaka University, Osaka 560-0043, Japan.}

\date{\today}

\draft

\begin{abstract}
Evolution of Fermi surface (FS) states of NdFeAs$_{1-x}$P$_x$O$_{0.9}$F$_{0.1}$ single crystals with As/P substitution has been investigated. 
The critical temperature $T_{\rm c}$ and the power law exponent ($n$) of temperature-dependent resistivity ($\rho(T) = \rho_0 + AT^n$) show a clear correlation above $x=$0.2, suggesting that $T_{\rm c}$ is enhanced with increasing bosonic fluctuation in the same type of FS state. 
Around $x=$0.2, all the transport properties show anomalies, indicating that $x$$\sim$0.2 is the critical composition of drastic FS change. 
The angle resolved photoemission spectroscopy has more directly revealed the distinct change of FS around $x=$0.2, that one hole FS disappears at Brillouin zone center and the other FS with propeller like shape appears at zone corner with decreasing $x$. 
These results are indicative of the existence of two types of FS state with different nesting conditions that are related with two $T_{\rm c}$-rising mechanisms in this system.

\end{abstract}

\pacs{74.70.Xa, 74.62.Dh, 74.25.F-, 74.25.Jb}

\maketitle
Since the discovery of iron-based superconductors (IBSs), many experimental and theoretical studies have been performed~\cite{Kamihara,Hosono}, but the superconducting (SC) mechanism of this system has not been clarified yet. 
The IBSs are multi-band systems with several 2-dimensional Fermi surfaces (FSs). 
It has been theoretically predicted that spin or orbital fluctuations play an important role in pairing mechanism, but this issue is still under debate~\cite{Mazin,Kuroki,Kontani}. 
Experimentally the empirical relation between $T_{\rm c}$ and the local structural parameter has been reported~\cite{Mizuguchi,Lee}. 
This was explained by the theory of K. Kuroki $et$ $al$. where the local structural parameters have close relation to the electronic band structure and FS states~\cite{Kuroki,FS}. 
However, there is no distinct experimental evidence for their relation. 
In addition, the FS topology of IBSs has a strongly material dependence~\cite{Lu1,Lu2,Nishi,Charnukha,Evtushinsky,Malaeb,Yoshida,Suzuki,Borisenko}, which causes the difficulty to clarify the SC mechanism of IBSs from the viewpoints of FS topology and related nesting conditions. 
Therefore, to understand the SC mechanism of IBSs, it is necessary to study experimentally the change of FS states related with nesting condition and pairing mechanism in adequate systems, where the local structural parameters and FS states can be systematically controlled. 

$R$FeAs$_{1-x}$P$_x$O$_{1-y}$F$_y$ ($R=$La or rare earth element) is good platform for such a study. 
There are several advantages. 
First, the isovalent As/P substitution induces a large and systematic change of crystal structure including the local Fe(As/P)$_4$ tetrahedron without any modification of band filling~\cite{Miyasaka1,Lai1,Miyasaka2,Lai2,Takemori1}. 
The deformation of local Fe(As/P)$_4$ tetrahedron is expected to seriously affect FS topology~\cite{Kuroki}. 
Second, we can cover a wide range of $T_{\rm c}$ from $\sim$0 K to $\sim$50 K using several $R$-systems~\cite{Miyasaka1}. 
Third, the strength of antiferromagnetic (AF) correlation can be systematically changed by varying $x$~\cite{Shiota}. 

In the previous studies, we have reported the transport properties in the polycrystalline samples of the 1111-type $R$FeAs$_{1-x}$P$_x$O$_{1-y}$(F/H)$_y$ ($R=$La, Pr and Nd)~\cite{Miyasaka1,Lai1,Miyasaka2}. 
In the La-1111 system, three SC regions exist near three different AF phases. 
In the low electron doping region of $y$$\sim$0.1, these systems show the crossover behavior between two different SC states by As/P substitution. 
Our investigation has indicated that $R$FeAs$_{1-x}$P$_x$O$_{1-y}$F$_y$ with F concentration of $y$$\sim$0.1 has two different $T_{\rm c}$-rising mechanisms, related to two different FS states and nesting conditions around $x=$0 and above $x=$0.2$\sim$0.4~\cite{Miyasaka1,Lai1,Miyasaka2,Shiota}. 
In the present work, to clarify the $x$-dependent change of electronic structures and FS topology directly, we have investigated the transport properties and the angle resolved photoemission spectroscopy (ARPES) using single crystals of NdFeAs$_{1-x}$P$_x$O$_{0.9}$F$_{0.1}$. 

The single crystals of NdFeAs$_{1-x}$P$_x$O$_{0.9}$F$_{0.1}$ with $x=$0-1.0 were grown from a mixture of NdAs, NdP, Fe$_2$O$_3$, Fe and FeF$_2$ under high pressure of 3.8 GPa by using cubic anvil press machine. 
It was heated up to $\sim$1400 $^{\circ}$C, kept for 12 hours and then slowly cooled down to $\sim$800 $^{\circ}$C for 72 hours. 
A typical size of single crystal was about 1$\times$1$\times$0.03 mm$^3$~\cite{Takemori2}. 
In this paper, we use the nominal P and F concentration as $x$ and $y$($=$0.1). 
The actual F concentration estimated by electron probe microanalyzer was approximately 0.05 in the whole $x$ region. 
The P content $x$ determined by energy dispersive X-ray spectroscopy was nearly the same as the nominal one. 

The magnetic susceptibility was measured in a magnetic field of 10 Oe, and the samples with $x \leqq $0.8 show a diamagnetic behavior below $T_{\rm c}$. 
The temperature ($T$) dependence of in-plane electrical resistivity $\rho(T)$ was measured by a standard four-probe method. 
The Hall coefficient $R_H$ was measured in magnetic fields parallel to $c$-axis up to 7 T. 
The ARPES measurements were carried out at BL-5U and 7U of UVSOR Facility at Institute for Molecular Science~\cite{Kimura}. 
The $k_z$ dependence of FS topology has been investigated by the ARPES with various incident photon energies. 
It has been confirmed that the topology of the observed FSs has been almost unchanged along $k_z$ direction. 
In this work, the in-plane ARPES measurements have been performed using incident photons with $h \nu=$36 eV polarized linearly. 
The MBS A-1 analyzer was used with the energy resolution of $\sim$15 meV. 
The angular resolution was $\sim$0.17$^{\circ}$. 
The single crystals were cleaved at $\sim$12 K in an ultrahigh vacuum ($\sim$1$\times$10$^{-10}$ Torr). 
The Fermi level ($E_F$) was calibrated using an evaporated gold film.

\begin{figure}[htp]
\begin{center}
\includegraphics[width=0.5\textwidth]{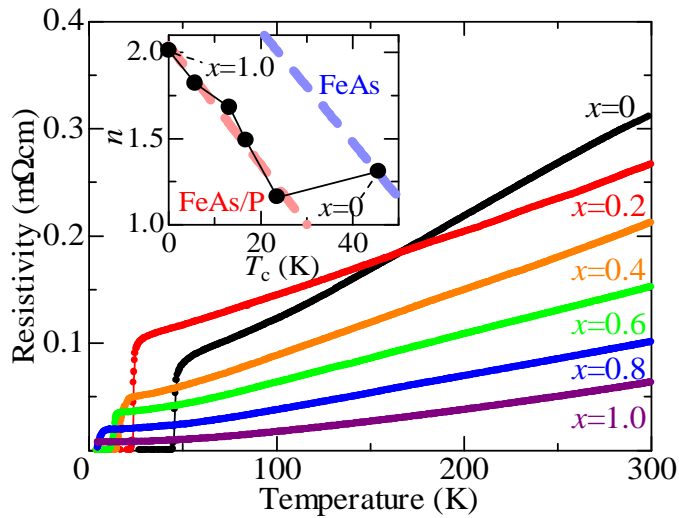}\\
\end{center}
\caption{(Color online) $T$ dependence of in-plane resistivity $\rho(T)$ for NdFeAs$_{1-x}$P$_x$O$_{0.9}$F$_{0.1}$ single crystals with $x=$0-1.0. Inset shows the relation between $T_{\rm c}$ and the power $n$ of $T$ in $\rho(T)$. The red and blue broken lines represent the two different correlations between $T_{\rm c}$ and $n$ in $R$FeAs$_{1-x}$P$_x$O$_{0.9}$F$_{0.1}$ ($x>$0.2-0.4) and $R$FeAsO$_{1-y}$F$_y$/$R$FeAsO$_{1-y}$ ($R=$La, Pr, and Nd)~\cite{Miyasaka1}.
}
\label{fig1}
\end{figure}

Figure 1 shows $\rho(T)$ for single crystals of NdFeAs$_{1-x}$P$_x$O$_{0.9}$F$_{0.1}$ with $x=$0-1.0. 
The $T$- and $x$-dependence of in-plane resistivity for single crystals are very similar to those of polycrystalline samples~\cite{Miyasaka1}. 
The room $T$ resistivity decreases with $x$. 
The samples below $x=$0.8 shows the SC transition, and $T_{\rm c}$ determined as zero resistivity $T$ is almost the same as the onset $T$ of the diamagnetism of magnetic susceptibility. 
On the other hand, $\rho(T)$ show no anomaly due to SDW transition in the whole $x$ region. 
$\rho(T)$ can be expressed as $\rho(T)=\rho_0+AT^n$ at low $T$s, where $\rho_0$ is residual resistivity, $n$ the power law exponent of $T$ and $A$ the coefficient. 
The fitting of $\rho(T)$ was performed between the onset $T$ of resistive transition and 100 K. 
The $x$ dependence of $T_{\rm c}$, $n$, $\rho_0$, and $A$ are presented in Fig. 2. 
NdFePO$_{0.9}$F$_{0.1}$ ($x=$1.0) undergoes no SC transition and $\rho(T)$ shows $T^2$ behavior, indicating that the system is a conventional Fermi liquid without any bosonic fluctuation. 
With decreasing $x$, $T_{\rm c}$ gradually increases up to $\sim$25 K, and $n$ systematically decreases and becomes close to unity at $x$$\sim$0.2. 
Below $x=$0.2, $T_{\rm c}$ is rapidly enhanced and $n$ is slightly increased. 

\begin{figure}[htp]
\begin{center}
\includegraphics[width=0.5\textwidth]{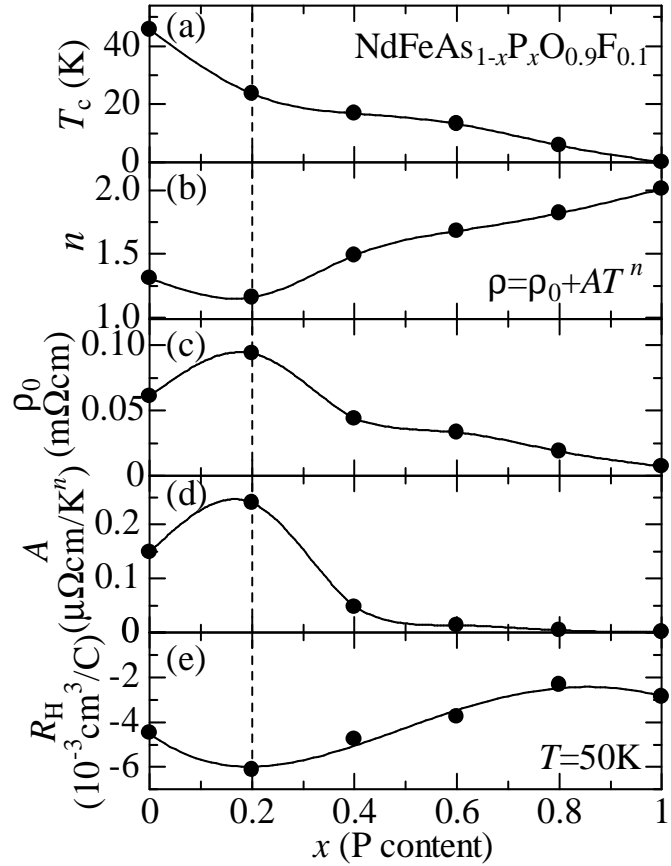}\\
\end{center}
\caption{$x$ dependence of physical properties for NdFeAs$_{1-x}$P$_x$O$_{0.9}$F$_{0.1}$ single crystals. (a) $T_{\rm c}$. (b) Power $n$ of $T$ in $\rho(T) = \rho_0+AT^n$. (c) Residual resistivity $\rho_0$. (d) The coefficient $A$. (e) $R_H$ at 50K. Solid lines are the guide for eyes. Broken vertical line indicates the critical concentration of $x=$0.2.}
\label{fig2}
\end{figure}

The $x$-dependence of $T_{\rm c}$ and $n$ has a clear relation in the composition region between $x=$0.2 and 1.0. 
The inset of Fig. 1 represents $T_{\rm c}$ vs. $n$ plot for the present single crystals. 
The figure also shows the two $T_{\rm c}$-$n$ relationships, which have been clarified by the previous study in polycrystalline samples~\cite{Miyasaka1,Ishida}. 
Above $x=$0.2, $T_{\rm c}$ and $n$ follow one of the linear $T_{\rm c}$-$n$ relationship, as shown by the red broken line in the inset of Fig. 1. 
This result indicates that the nesting between hole- and electron-FSs becomes better with decreasing $x$, which gives stronger bosonic fluctuation as monitored by the power $n$, and results in the enhancement of $T_{\rm c}$. 
On the other hand, the data point of $T_{\rm c}$ and $n$ for NdFeAsO$_{0.9}$F$_{0.1}$ ($x=$0) is located on the other $T_{\rm c}$-$n$ scaling line. 
This reveals that the FS topology and the nesting conditions are quite different between the composition regions at $x$$<$0.2 and $x$$>$0.2, and $x$$\sim$0.2 is the critical concentration for the change of FS as well as related bosonic fluctuations. 
As shown in Fig. 2(e), the $x$-dependent $R_H$ at 50 K shows a broad minimum around $x=$0.2, suggesting the critical change of electronic structure too. 

Recent experimental investigations of transports and NMR have indicated that the LaFeAs$_{1-x}$P$_x$O has two different AF states around $x=$0 and 0.6, suggesting the change of FS state and related nesting condition by As/P solid solution~\cite{Lai1,Mukuda1,Kitagawa}. 
These AF orders are suppressed by F doping, and $T_{\rm c}$ shows a double-peak structure at $x=$0 and 0.6 in LaFeAs$_{1-x}$P$_x$O$_{1-y}$F$_y$ with $y=$0.05~\cite{Lai1}. 
With further F substitution, these peaks of $T_{\rm c}$ merge and create a single peak around $x=$0.4 in the $y=$0.1 system~\cite{Miyasaka1,Miyasaka2}. 
In LaFeAs$_{1-x}$P$_x$O$_{1-y}$F$_y$ with $y=$0.05 and 0.1, $n$ of $\rho(T)$ is systematically decreased from 2 to about 1, when $x$ decreases down to $x$$\sim$0.6 and 0.4, respectively. 
These results are suggestive of the enhancement of bosonic fluctuation such as AF fluctuation. 
NMR studies in LaFeAs$_{1-x}$P$_x$O$_{1-y}$F$_y$ have revealed that $(T_1T)^{-1}$ shows a distinct enhancement around ($x$, $y$)$=$(0.6, 0.05) and (0.4, 0.1), that also indicates the existence of strong AF fluctuation~\cite{Shiota,Mukuda2}. 
In the present NdFeAs$_{1-x}$P$_x$O$_{0.9}$F$_{0.1}$, the power $n$ of $\rho(T)$ also shows similar behavior to La-system, indicating the existence of strong AF fluctuation around $x=$0.2. 

As shown in Fig. 2, $\rho_0$ and $A$ are clearly enhanced around $x=$0.2. 
These behaviors cannot be explained by the conventional spin fluctuation theory, but they suggest the existence of additional bosonic fluctuations. 
The similar anomalies together with $T$-linear $\rho(T)$ were observed near the quantum critical point (QCP) in heavy fermion Ce compounds~\cite{Yuan1,Yuan2,Seyfarth,Watanabe}. 
These heavy fermion systems have maximum $T_{\rm c}$ values and show anomalies of $\rho_0$ and $A$ at critical pressure ($P$) of valence transition of Ce ion, which is apart from magnetic QCP in $P$-$T$ phase diagram. 
In NdFeAs$_{1-x}$P$_x$O$_{0.9}$F$_{0.1}$, the isovalent As/P substitution does not cause the change of Fe 3$d$ band filling, but induces the distinct change of electronic structure near $E_F$ as described later, and concomitantly the charge fluctuation at critical concentration $x$$\sim$0.2, which could enhance $T_{\rm c}$ at $x$$\sim$0.2.

\begin{figure}[htp]
\begin{center}
\includegraphics[width=0.8\textwidth]{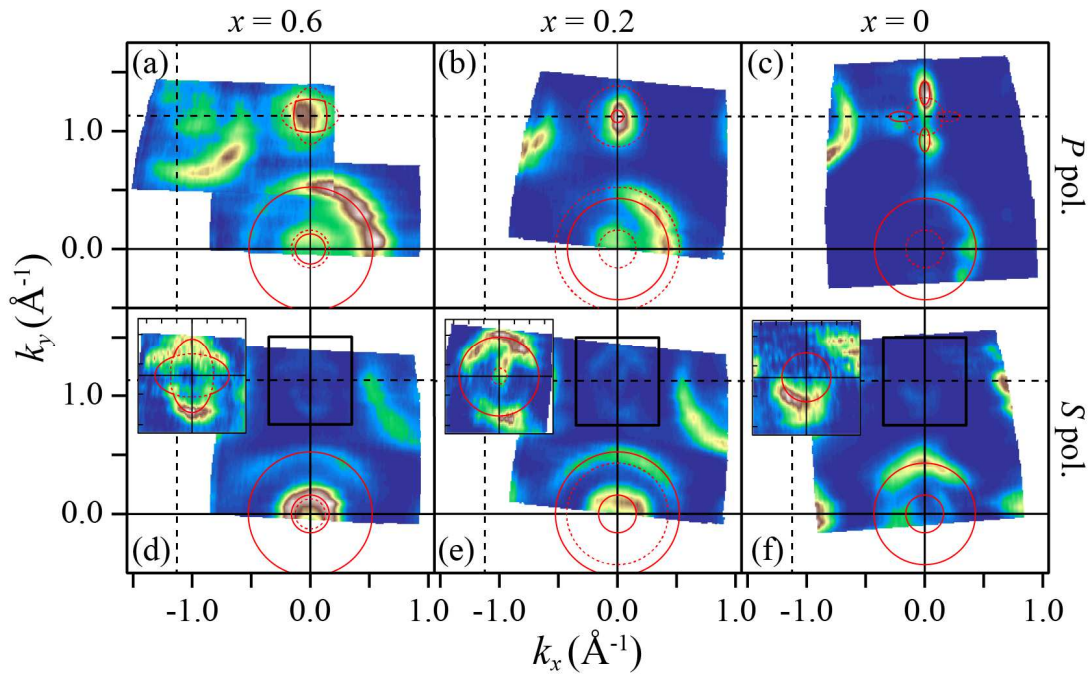}\\
\end{center}
\caption{(Color online) In-plane FS mapping of NdFeAs$_{1-x}$P$_x$O$_{0.9}$F$_{0.1}$ single crystals with $x=$0.6 ((a),(d)), 0.2 ((b),(e)), and 0 ((c),(f)), respectively. Upper (a)-(c) and lower (d)-(f) panels show the results in $P$ and $S$ geometries, respectively. Insets of (d)-(f) show the results around BZ corner by using different color gradation. The solid lines indicate the FSs observed in the polarization geometry, while the broken ones are the FSs observed in the other geometry.}
\label{fig3}
\end{figure}

This change of electronic structure at the critical concentration of $x$$\sim$0.2 has been directly observed by ARPES measurement. 
Figure 3 shows the in-plane FS mapping for $x=$0.6, 0.2 and 0 at $\sim$12 K. 
Here, $x$ and $y$ axes are parallel to the Fe-Fe bond direction within the plane. 
In the $P$ ($S$) polarization geometry, the electric field vector of incident photons is parallel (perpendicular) to the mirror plane defined by the analyzer silt parallel to $x$ axis and the normal vector ($z$ axis) of cleaved sample surface. 
The propagator vector of the incident photons are located in the mirror plane ($xz$ plane)~\cite{Kimura}. 
In the present ARPES measurements, $x$ axis is not perfectly parallel to the analyzer slit, but the misalignment angel is less than $\sim$6 $^{\circ}$. 
In Fig. 3, the photoemission intensity has been integrated over the energy range of $E_F \pm$5 meV. 

The FSs with $xz$, 3$z^2-r^2$ and $x^2-y^2$ characters of Fe 3$d$ orbital can be detected in the $P$ polarization geometry, while those with $yz$ and $xy$ in the $S$ polarization around Brillouin zone (BZ) center~\cite{Zhang1,Zhang2}. 
First, we focus our discussion on the data for the BZ center. 
As is seen in Figs. 3(a) and (d), the two small FSs observed in the $P$ and $S$ polarization geometries have a quite similar size, indicating that the bands related with these FSs are almost degenerated near the BZ center (0, 0) for $x=$0.6. 
Therefore, the orbital characters of these FSs have been assigned to $xz$ and $yz$. 
This assignment is consistent with the previous theoretical and experimental studies in the related iron-phosphorous material, LaFePO~\cite{Kuroki,Lu1,Lu2,Nishi,Vildosola}. 
In the 1111 system, there are theoretically predicted other electronic bands with $xy$ and 3$z^2-r^2$ characters near BZ center~\cite{Kuroki,Vildosola}. 
The outer FS observed in the $P$ polarization geometry may have 3$z^2-r^2$ character. 
As seen in Figs. 3(a) and 4(a), however, the FS size is larger and the energy level of the related band is higher than those predicted by the theoretical studies~\cite{Kuroki,Vildosola}. 
The orbital character of outer FS is unclear at present. 
In addition, the previous ARPES studies indicated that the 1111-type compounds suffer from the polar surface problem~\cite{Charnukha,Zhang3}. 
Due to the lack of a charge neutral cleavage plane, the doping level of the cleaved surface is significantly different from the nominal doping level expected from the chemical composition. 
This outer FS might be originated from the surface electronic state. 
Therefore, we discuss here only the inner FSs, which show a systematic change with the P content ($x$) suggesting the change of the bulk electronic state. 

As shown in Fig. 3, the large and small FSs were observed around BZ center also in the $x=$0.2 and 0 samples. 
The small FS with $xz$ character observed in the $P$ polarization geometry disappears below $x=$0.2 (Figs. 3(a)-(c)), while the FS with $yz$ character in the $S$ geometry are almost unchanged with the decrease of $x$ (Figs. 3(d)-(f)). 
The disappearance of the $xz$-FS below $x=$0.2 can be seen also in the band dispersion. 
Figure 4 illustrates the ARPES intensity mapping that corresponds to the band dispersion on the $P$ polarization geometry for $x=$0.6, 0.2 and 0. 
As shown in Figs. 4(a)-(c), the energy level of inner $xz$ band in the $P$ geometry decreases with decreasing $x$. 
The top of $xz$ band touches $E_F$ at $x$$\sim$0.2 (Fig. 4(b)), and this band sinks down perfectly below $E_F$ at $x$$\sim$0. 
This $x$ dependence of the bands can be also seen in the energy distribution curves (EDCs) in Fig. 4(d). 
Resultantly, the $xz$ FS around BZ center disappears below $x$$\sim$0.2, indicating the change of FS states and the related nesting condition at this critical $x$. 

The previous ARPES studies have indicated that LaFePO has double-degenerated small hole FSs around BZ center, while the 1111-type FeAs compounds have only one small FS~\cite{Lu1,Lu2,Nishi,Charnukha}. 
The present study has clarified the missing link of FS state between 1111-type FeP and FeAs compounds. 
In this system, As/P substitution induces the change of electronic structure concomitantly with the monotonic transformation of the local structural parameters around Fe ion~\cite{Miyasaka2,Lai2,Takemori1}. 
Particularly, the bond angle of As/P-Fe-As/P shows the linear $x$-dependence from $\sim$111$^{\circ}$ for $x=$0 to $\sim$118$^{\circ}$ for $x=$1.0 in NdFeAs$_{1-x}$P$_x$O$_{0.9}$F$_{0.1}$, which have been determined by our structural analysis for the polycrystalline samples using the synchrotron X-ray diffraction. 
The present ARPES results clearly demonstrate that the systematic change of local structure directly affects the FS state and band structure near BZ center, particularly the $xz$ hole band and its FS. 
The band calculation, where the As-Fe-As bond angle is virtually changed, has demonstrated a similar FS change and the $xz$/$yz$ band splitting near $\Gamma$ point with decreasing bond angle down to $\sim$109$^{\circ}$~\cite{Usui1,Usui2}. 
According to the calculation, the $xz$/$yz$ band splitting is accompanied by the lifting of energy level of $xy$ band up to $E_F$, which has been observed in our preliminary ARPES experiments. 

\begin{figure}[htp]
\begin{center}
\includegraphics[width=0.8\textwidth]{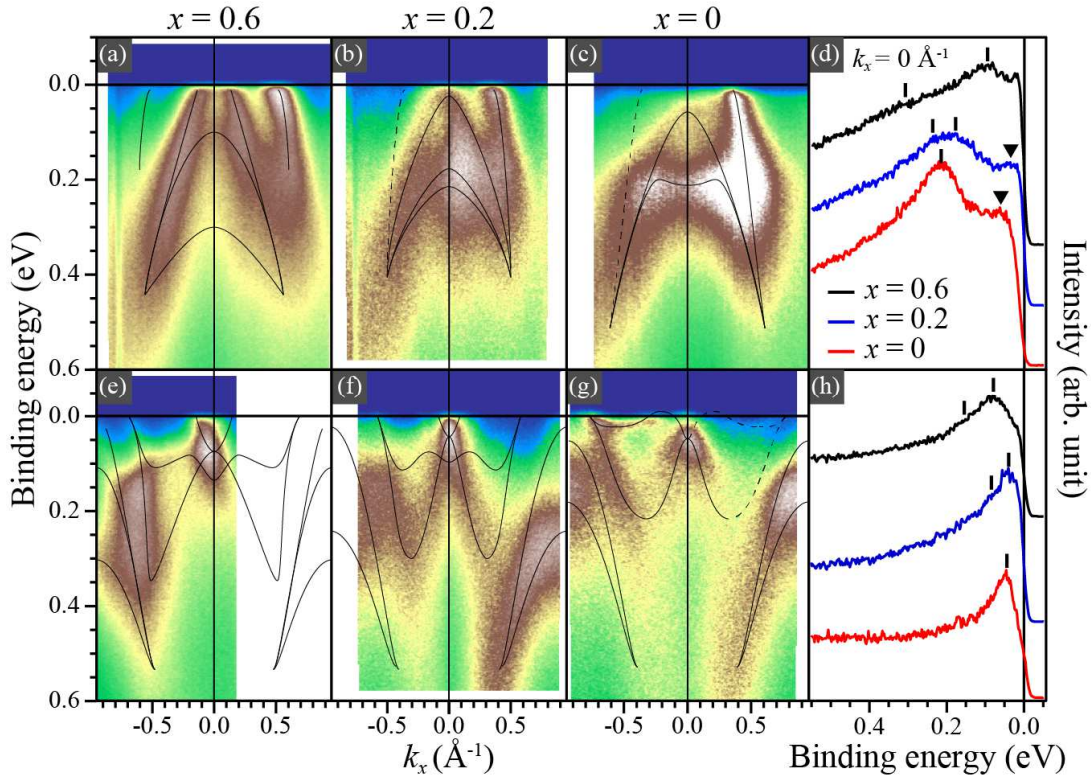}\\
\end{center}
\caption{(Color online) ARPES intensity mapping in the $P$ polarization geometry for NdFeAs$_{1-x}$P$_x$O$_{0.9}$F$_{0.1}$ single crystals with $x=$0.6 ((a),(e)), 0.2 ((b),(f)), and 0 ((c),(g)). Panels (a)-(c) and (e)-(g) show the results of energy-momentum cut around BZ center (0, 0) and around corner (0, $\pi$), respectively. 
Solid lines indicated the band dispersion estimated from the anomalies such as peaks and kinks in EDCs at various $k_x$ and in momentum distribution curves (MDCs) at various binding energies. The band dispersions should be symmetric with respect to the line at $k_x=$0 because of the crystal symmetry, but some parts of the band dispersion indicated by the broken lines cannot be estimated from the EDCs and MDCs due to the low ARPES intensity. In order to draw these band dispersions, the several previous works have been referred~\cite{Lu1,Nishi,Charnukha,Zhang2}. 
Panels (d) and (h) show the EDCs of $x=$0, 0.2 and 0.6 at BZ center and corner ($k_x=$0), respectively. The bars indicate the energy position of top or the bottom of the bands shown in panels (a)-(c) and (e)-(g). In panel (d), the triangles indicate the energy position of top of the $xz$ band.}
\label{fig4}
\end{figure}

Next, we see the FSs around the BZ corner (0, $\pi$) in Fig. 3. 
Since the photoemission intensity near $E_F$ is low around BZ corner in the $S$ polarization geometry, the results are presented by different color gradation in the insets of Figs. 3(d)-(f). 
At $x=$0.6, two FSs with elliptical shape exist around BZ corner. 
(Figs. 3(a) and (d)) 
The FS shape around BZ corner drastically changes with decreasing $x$. 
At $x=$0.2, there are two cylindrical FSs, while the FS for $x=$0 has quite different shapes, namely one is cylindrical and the other is propeller like. 

As shown in Fig. 4, the band dispersion around BZ corner is complicated and its $x$ dependence is non-monotonic. 
In the $P$ geometry of $x=$0.6 and 0.2 samples, one electron band is observed around BZ corner, and it creates the elliptical and cylindrical FSs, respectively. 
On the other hand, there are two other bands with complicated dispersions around the binding energy range of 0$\sim$0.3 eV. 
As shown in Figs. 4(e)-(h), all the bands tend to lift up with decreasing $x$. 
For $x=$0, the bottom of the electron band with parabolic dispersion shifts up above $E_F$ and this electron FS disappears. 
One of the other bands, located around the binding energy of $\sim$0.1 eV for $x=$0.6 and 0.2, shows the relatively large energy shift between $x=$0.2 and 0. 
(Fig. 4(h)) 
At $x=$0, this band crosses $E_F$ and makes the propeller like hole FSs, which have been observed in the other 1111-type pure As systems~\cite{Nishi,Charnukha}. 
The surface electronic state may affect some of the FSs and the bands around BZ corner~\cite{Charnukha,Zhang3}, but the observed $x$ dependence is essentially originated from the change of bulk electronic state. 

The discontinuous change of band dispersion and FSs between $x=$0.2 and 0 induces the change of nesting condition and the different bosonic fluctuations. 
As a result, it causes the different $T_{\rm c}$-rising mechanism below and above $x$$\sim$0.2. 
Above $x$$\sim$0.2, 2 cylindrical hole FSs around BZ center and 2 electronic FSs with almost cylindrical shape around zone corner have a good nesting condition, which produces the strong spin fluctuation. 
The good FS nesting in the P-doping region has been theoretically predicted by H. Usui $et$ $al$.~\cite{Usui3}, and the enhancement of the low-energy spin fluctuation at $x$$\sim$0.2 has been observed by the present study of $\rho(T)$ and also by the previous NMR study in La-1111 systems~\cite{Shiota}. 
This spin fluctuation regime does not remain below $x$$\sim$0.2 because one hole FS disappears and the FS topology around BZ corner becomes complicated, giving the moderate nesting condition. 
At $x$$\sim$0, the $xy$ hole band exists very near $E_F$ around BZ center, which suggests some other types of bosonic fluctuation. 
First, this band and FSs around BZ corner may induce the spin fluctuation with relatively higher energy. 
In fact, the previous results of NMR study suggested the existence of the spin fluctuation with higher energy~\cite{Shiota}. 
Second, the $xy$ band near $E_F$ may also induce some charge fluctuation. 
These spin or charge fluctuation could be another candidate of bosonic fluctuation related with the SC mechanism in the 1111-type FeAs system. 

In summary, the transport and ARPES studies have been performed using NdFeAs$_{1-x}$P$_x$O$_{0.9}$F$_{0.1}$ single crystals to clarify the $x$ dependence of bosonic fluctuation and electronic bands that should be related with the SC mechanism. 
In the present system, $T_{\rm c}$ gradually increases from 0 K to $\sim$25 K with decreasing $x$ from $x=$1.0 to 0.2, and is rapidly enhanced below $x=$0.2. 
At 0.2$\leqq x \leqq$1.0, $T_{\rm c}$ increases and the $T$ dependence of $\rho(T)$ changes from $T^2$ to $T$-linear behavior monotonously with decreasing $x$. 
This close correlation of $T_{\rm c}$ and $\rho(T)$ implies that $T_{\rm c}$ is enhanced with increasing bosonic fluctuation with the same FS topology. 
On the other hand, $T_{\rm c}$ and the $T$ dependence of $\rho(T)$ have a different relationship below $x$$\sim$0.2. 
The slope of $\rho(T)$, the residual resistivity and $R_H$ have anomalies around $x=$0.2. 
These results indicate that the FS and related bosonic fluctuations drastically change at $x$$\sim$0.2, which is consistent with the ARPES observation of a clear change of FS around $x=$0.2. 
The hole FS with $xz$ orbital character around BZ center disappears at $x$$\sim$0.2 with decreasing $x$. 
Around BZ corner, the FS shape is clearly changed between $x=$0.2 and 0. 
The significant variation of FSs around $x=$0.2 must induce the change of bosonic fluctuations and resultantly cause a further enhancement of $T_{\rm c}$ below  $x$$\sim$0.2 in As/P solid solution 1111 systems. 

\begin{acknowledgments}
We thank K. Kuroki, H. Usui and H. Mukuda for helpful discussion. 
This work was supported by Grants-in-Aid for Scientific Research from MEXT and JSPS, and by the JST project (TRIP and IRON-SEA) in Japan. 
Part of this work was supported by the Use-of-UVSOR Facility Program (BL5U and 7U, 2014-2017) of the Institute for Molecular Science.

\end{acknowledgments}




\begin{thebibliography}{99}
\bibitem{FS}In this paper, ``FS states" means electronic states with characteristic FSs observed in IBSs. 
For example, the theoretical study by K. Kuroki $et$ $al$. at the early stage has predicted that one of 1111-type IBSs, NdFeAsO system has several 2-dimensional hole and electron FSs including the $xy$ hole FS, while the $xy$ hole FS disappears in the related material LaFePO~\cite{Kuroki}. 
In other words, their investigation has theoretically demonstrated that these 1111-type systems have different FS states.
\end{thebibliography}
\end{document}